\documentclass{JHEP3}
\usepackage{graphicx}
\usepackage{amssymb}
\usepackage[mathscr]{eucal}
\usepackage{amsmath}
\usepackage{cite}
\usepackage{afterpage}
\usepackage{slashed}
\usepackage{feynmp}

\newcommand{\gsimm}{\raise.3ex\hbox{$>$\kern-.75em\lower1ex\hbox{$\sim$}}}
\newcommand{\lsimm}{\raise.3ex\hbox{$<$\kern-.75em\lower1ex\hbox{$\sim$}}}

\newcommand{\be}{\begin{equation}}
\newcommand{\ee}{\end{equation}}
\newcommand{\ba}{\begin{eqnarray}}
\newcommand{\ea}{\end{eqnarray}}

\newcommand{\bea}{\begin{eqnarray*}}
\newcommand{\eea}{\end{eqnarray*}}

\title{Casimir, Gravitational and Neutron Tests of Dark Energy}

\author{Philippe Brax\\
  Institut de Physique Th\'eorique, CEA, IPhT, CNRS, URA 2306,
  F-91191Gif/Yvette Cedex, France \\ E-mail:
  \email{philippe.brax@cea.fr}}

\author{Anne-Christine Davis\\
  DAMTP, Centre for Mathematical Sciences, University of Cambridge,
  CB3 0WA, UK\\E-mail:
  \email{A.C.Davis@damtp.cam.ac.uk}}

\date{today}
\abstract{We investigate laboratory tests of dark energy theories which modify gravity in a way  generalising the inverse power law chameleon models. We make use of the tomographic description of such theories which captures  $f(R)$ models in the large curvature limit, the dilaton and the symmetron.  We  consider their
effects in various experiments where the presence of a new scalar interaction may be uncovered. More precisely, we focus on the Casimir,  Eot-wash  and neutron experiments. We show that dilatons,  symmetrons and generalised chameleon models are efficiently testable in the laboratory. For generalised chameleons, we revise  their status in the light of forthcoming Casimir experiments like CANNEX in Amsterdam and show that they are within reach of detection.  }

\begin{document}
\section{Introduction}

Laboratory tests \cite{Adelberger:2003zx,Lamoreaux:1996wh,Nesvizhevsky:2003ww} offer a complementary approach to astrophysical and cosmological observations for dark energy/modified gravity theories \cite{Copeland:2006wr,Clifton:2011jh} involving one scalar field coupled to matter in a conformal way \cite{Jain:2013wgs}.  Astrophysical and cosmological probes are sensitive to a deviation of the equation of state for the dark energy fluid from the  standard $\Lambda$-CDM model  and its impact on the background cosmology. They are also affected  by new scalar interactions in the Mpc range  and its effects on structure formation at the perturbative level \cite{Lombriser:2014dua}. Developments in N-body simulations  to study non-linear effects in the growth of structures \cite{Li:2011vk} combined with new surveys like Euclid \cite{Amendola:2012ys} will give constraints on the large scale properties of dark energy and modified gravity scenarios which may reach and improve on the level already attained by solar system tests\cite{Bertotti:2003rm,Williams:2012nc}. They will complement them in a range of scales where gravitational properties have never been extensively studied. On the theoretical side, the most general type of scalar models where both dark energy and modifications of gravity can be envisaged has been recently rediscovered in the form of the Horndeski theories\cite{Horndeski:1974wa,Deffayet:2011gz}. Similarly, bimetric theories of gravity \cite{Hassan:2014gta} allow one to consider theoretically sound extensions of the original Pauli-Fierz theory of massive gravity. For all these models, the landscape of their possible physical consequences has only been explored in some corners where both linear on very large scales and non-linear effects,  on small scales down to laboratory ones,  can be analysed.
In this paper, we will present new results for a subset of scalar field models where the effects of the scalar field are screened by either the chameleon \cite{Khoury:2003aq,KhouryWeltman} or the Damour-Polyakov \cite{Damour:1994zq} mechanisms on small scales.
These tomographic theories \cite{Brax:2011aw,Brax:2012gr} generalise the inverse power law chameleons, the $f(R)$ theories \cite{Hu:2007nk}, dilatons \cite{Brax:2010gi} and symmetrons \cite{Pietroni:2005pv,Hinterbichler:2010es}.
We consider laboratory experiments which are searching for deviations from the electromagnetic Casimir pressure \cite{Lamoreaux:1996wh}, testing the existence  of extra forces like Eot-wash \cite{Adelberger:2003zx} and measuring the neutron energy levels in the terrestrial gravitational field \cite{Nesvizhevsky:2003ww}. Extensions to atomic and neutron interferometry experiments \cite{Brax:2013cfa,Burrage:2014oza} can also be considered.   Using the tomographic method, we are able to express the Casimir pressure due to the scalar field, the torque between the plates of the Eot-wash experiment and the displacement of the neutron energy levels in a simple manner \cite{Brax:2011hb}. For inverse power law chameleons and symmetrons, our  analytical results are compatible with numerical simulations of the exact experimental setup \cite{Upadhye:2012qu,Upadhye:2012rc}. They can be applied to all models described tomographically.

 We also pay attention to the issue of quantum corrections \cite{Upadhye:2012vh} and calculate the one loop effects in a homogeneous medium. Such effects can be large in dense matter and can invalidate the predictions made at the classical level. This is particularly true of inverse power law chameleons where the quantum corrections in the boundary plates of laboratory experiments can be large for relatively low couplings to matter. For larger values of the coupling, one cannot guarantee that the scalar field profile between the plates is maintained as the boundary values for the scalar may have been altered drastically by quantum effects. Fortunately, for chameleons at large enough couplings the homogeneous solution does not hold anymore and the field forms bubbles at the atomic level \cite{Brax:2013cfa}. These bubbles are quantum stable when quantum corrections inside nuclei are tamed by imposing a vanishingly small coupling in nuclear matter. In this case, the bubble solution between the boundary plates is not sensitive to quantum corrections and experiments tackling the large coupling regime of
chameleons such as the next generation of Casimir experiments will give valuable information on the chameleon's parameter space .

We use our results for $f(R)$ models in the large curvature regime and find that their effects in the laboratory are negligible. For dilatons, we find that the 2006 Eot-wash measurements at a distance of $55 \ \mu$m give a strong restriction on the cosmological mass of the scalar field, although still two orders of magnitude below the bound from the tests of the equivalence principle by the Lunar Ranging experiment. Symmetrons with cosmological effects cannot be effectively tested in the laboratory although the ones with a phase transition in rather dense media can be. Finally, generalised chameleons with inverse power law potentials are found to be very close to being detectable by the next generation of Casimir experiments \cite{Brax:2010xx} as soon as their sensitivity will drop below one pN/${\rm cm^2}$.

In section 2, we introduce the tomographic models. In section 3, we consider planar field configurations. In section 4, we present the various experimental situations that we will consider and calculate their observables for various models in section 5. In section 6, we analyse  the quantum corrections for tomographic models. In section 7, we deduce the present laboratory constraints on tomographic models and the forecasts for inverse power law chameleons.
We conclude in section 8.

\section{Tomographic Models}

Inverse power law chameleon models and their generalisations are scalar-tensor theories described by the Lagrangian
\be
S=\int d^4 x \sqrt{-g}(\frac{R}{16\pi G_N} -\frac{(\partial \phi)^2}{2} -V(\phi))+S_m (\psi, A^2(\phi) g_{\mu\nu})
\label{act}
\ee
where $A(\phi)$ is an arbitrary function which specifies the coupling between matter fields $\psi$ and the scalar $\phi$. The coupling to matter itself  is given by
\be
\beta (\phi)= m_{\rm Pl} \frac{d \ln A(\phi)}{d \phi}.
\ee
The most important feature of these models  is that the scalar field dynamics are determined by an effective potential which takes into account the presence of the conserved matter density $\rho$ of the environment
\be
V_{\rm eff}(\phi) =V(\phi) +(A(\phi)-1) \rho.
\ee
When the effective potential acquires a matter dependent minimum $\phi_{}(\rho)$, for instance when $V(\phi)$ decreases and $A(\phi)$ increases,
the mass of the scalar at the minimum is also matter dependent $m(\rho)$.
Scalar-tensor theories whose effective potential $V_{\rm eff}(\phi)$ admits a density dependent minimum $\phi (\rho)$ can all be described
parametrically from the sole knowledge of the mass function $m(\rho)$ and the coupling $\beta (\rho)$ at the minimum of the potential \cite{Brax:2012gr,Brax:2011aw}. It is often simpler to characterise the functions $m(\rho)$ and $\beta(\rho)$ using the time evolution of the matter density of the Universe
\be
\rho(a)=\frac{\rho_0}{a^3}
\ee
where $a\le 1$ is the scale factor of the Universe whose value now is $a_0=1$. This allows one to describe characteristic models in a simple way, even in situations like laboratory tests where no cosmology is involved. The field value is given by
\be
\frac{\phi (a)-\phi_c}{m_{\rm Pl}}= 9\Omega_{m0} H_0^2 \int_{a_c}^{a} da \frac{\beta (a) }{a^4m^2(a)},
\label{tom}
\ee
where the Hubble rate now is $H_0\sim 10^{-43}$ GeV and the matter fraction is $\Omega_{m0}\sim 0.27$.
We have identified the mass as the second derivative
\be
m^2 (a)= \frac{d^2 V_{\rm eff}}{d\phi^2}\vert_{\phi=\phi (\rho(a))}
\ee
and the coupling
\be
\beta (a)= m_{\rm Pl} \frac{d\ln A}{d\phi}\vert_{\phi=\phi(\rho(a))}.
\ee
The potential value is given by
\be
V(a)-V_c= -27\Omega_{m0}^2 H_0^4 \int_{a_c}^{a} da \frac{\beta^2 (a) m_{\rm Pl}^2}{a^7m^2(a)}.
\ee
This parameterisation allows one to obtain $V(\phi)$ and $A(\phi)$ implicitly from $m(a)$ and $\beta (a)$.

\subsection{Inverse power law chameleons}

Chameleons with a potential of the type
\be
V(\phi)=\Lambda^4 +\frac{\Lambda^{4+n}}{\phi^n} + \dots
\ee
where $n>0$, $\Lambda\sim 10^{-3}$ eV is the cosmological vacuum energy now, and the coupling function is
\be
A(\phi)=\exp(\frac{\beta\phi}{m_{\rm Pl}}),
\ee
can be reconstructed using
\be
\beta(a)=\beta
\ee
and
\be
m(a)= m_0 a^{-r}
\ee
where $r= \frac{3(n+2)}{2(n+1)}$. The mass scale $m_0$ is determined by
\be
m_0^{2(n+1)}=\frac{(n+1)^{n+1}}{3n}\frac{(3\beta \Omega_{m0}H_0^2 m_{\rm Pl})^{n+2}}{ \Lambda^{4+n} }
\ee
which gives dimensionally
\be
m_0\sim \beta^{(n+2)/2(n+1)} (\frac{m_{\rm Pl}}{H_0})^{n/4(n+1)} H_0.
\ee
This implies that inverse chameleon models have a cosmological interaction range $1/m_0$  much shorter than the size of the observable Universe for $\beta_0\gtrsim 1$.

\subsection{Large curvature f(R)}
A large class of interesting models of the chameleon type consists of the large curvature $f(R)$ models with the action
\be
S=\int d^4x \sqrt{-g} \frac{f(R)}{16\pi G_N}
\ee
where the function $f(R)$ is expanded in the large curvature regime
\be
f(R)=\Lambda_0 + R -\frac{f_{R_0}}{n} \frac{R_0^{n+1}}{R^n}.
\ee
Here $\Lambda_0$ is the cosmological constant term necessary to lead to the late time acceleration of the Universe and $R_0$ is the present day curvature.  These models can be reconstructed using the constant $\beta(a)=1/\sqrt{6}$ and the mass function
\be
m(a)= m_0 (\frac{4\Omega_{\Lambda 0}+ \Omega_{m0} a^{-3}}{4\Omega_{\Lambda 0}+ \Omega_{m0}})^{(n+2)/2}
\ee
where the mass on large cosmological scale is given by
\be
m_0= H_0 \sqrt{\frac{4\Omega_{\Lambda 0}+ \Omega_{m0} }{(n+1) f_{R_0}}},
\ee
and $\Omega_{\Lambda 0} \approx 0.73$ is the dark energy fraction now \cite{Brax:2012gr}. When $a\ll 1$ corresponding to physical situations where the environment is dense, the mass dependence on $a$ is a power law
\be
m(a) \sim m_0 a^{-r}
\ee
where $r=\frac{3(n+2)}{2}$.

\subsection{Dilaton}

Another relevant example is the environmentally dependent dilaton \cite{Brax:2010gi}. This model is inspired by string theory in the large string coupling limit
with an exponentially runaway potential
\be
V(\phi)=V_0 e^{-\frac{\phi}{m_{\rm Pl}}}
\ee
where $V_0$ is determined to generate the acceleration of the Universe now and the coupling function is
\be
A(\phi)=\frac{A_2}{2m_{\rm Pl}^2} (\phi-\phi_\star)^2.
\ee
These models can be described using the coupling function in the matter dominated era
\be
\beta(a)= \beta_0 a^3
\ee
where $\beta_0$ is related to $V_0$ and is determined by requiring that $\phi$ plays the role of late time dark energy which sets $\beta_0=\frac{\Omega_{\Lambda 0}}{\Omega_{m0}}\sim 2.7$, and the mass function which reads
\be
m^2(a)= 3 A_2 \frac{H_0^2}{a^3}
\ee
and is proportional to the Hubble rate with the mass on cosmological scales now given by $m_0=\sqrt{3 A_2} H_0$.

\subsection{Symmetrons}

Another example is the symmetron where a scalar field has a quartic potential with a non-vanishing minimum
\be
V(\phi)= V_0 + \frac{\lambda}{4} \phi^4 -\frac{\mu^2}{2} \phi^2
\ee
and a coupling function
\be
A(\phi)= 1 + \frac{\beta_\star}{2\phi_\star m_{\rm Pl}} \phi^2
\ee
where the transition from the minimum of the effective potential  at the origin to a non-zero value happens at $a=a_\star$. This is a second order phase transition where the mass vanishes.
Defining
\be
m_\star= \sqrt 2 \mu,\ \phi_\star= \frac{2\beta_\star \rho_\star}{m_\star^2 m_{\rm Pl}},\ \lambda= \frac{\mu^2}{\phi_\star^2}
\ee
where $\rho_\star=\frac{\rho_{m0}}{a_\star^3}$,
the model can be reconstructed using
\be
m(a)=m_\star \sqrt{ 1-(\frac{a_\star}{a})^3}
\ee
and
\be
\beta(a)=\beta_\star \sqrt{1-(\frac{a_\star}{a})^3}
\ee
for $a>a_\star$ and $\beta (a)=0$ for $a<a_\star$. In dense environment, the field is at the origin while in a sparser one with $a>a_\star$ we have
\be
\phi= \phi_\star \sqrt{1-(\frac{a_\star}{a})^3}.
\ee

\subsection{ Generalised power law models}

The inverse power law chameleons, the dilaton and f(R) models in a dense environment are all described by power law functions
\be
m(a)= m_0 a^{-r}, \ \beta(a)=\beta_0 a^{-s}
\ee
for different choices of $r$ and $s$. In fact, all these models can be defined by a potential
\be
V(\phi) = V_0 +\epsilon \Lambda^{4-p}\phi^p
\ee
 where $V_0$ is an arbitrary constant, and
\be
p=\frac{2r-6-2s}{2r-3-s}
\ee
as long as $(2r-3-s)>0$. The sign $\epsilon=\pm 1$ is positive when $p<0$ and vice versa. As for inverse power law chameleon models, it is convenient to  introduce the effective scale
\be
\Lambda^{4-p}=\frac{27}{\vert 2r-6-2s\vert } \frac{\Omega_{m0}^2 \beta_0^2 H_0^4 m_{\rm Pl}^2}{m_0^2} (\frac{2r-3-s}{9} \frac{m_0^2}{ \Omega_{m0} \beta_0 H_0^2 m_{\rm pl}})^p
\ee
which is a function of both $m_0$ and $\beta_0$. For inverse power law chameleons, it is taken to be the dark energy scale.
The coupling function becomes
 \be
A(\phi)= \frac{\beta_0}{m_{\rm Pl}} \frac{\phi^l}{M^{l-1}}
\ee
where the power $l$ is given by
\be
l=\frac{2r-3-2s}{2r-3-s}
\ee
 and the coupling scale is
\be
M^{1-l}= \frac{\Omega_{m0}}{l} (\frac{9}{2r-3-s}\frac{\Omega_{m0}\beta_0 H_0^2 m_{\rm pl}}{m_0^2})^{\frac{s}{2r-3-s}}.
\ee
Although very explicit, this field parameterisation of the models is cumbersome. We will mostly use the $(m(a),\beta (a))$ definition in the following.

\section{Planar Solutions in Modified Gravity}

\subsection{Planar Configurations}

The experimental setups that we will consider in the following can all be well approximated by two infinite plates separated by a distance $2d$. In this case, the scalar field satifies the Klein-Gordon equation which reduces to
\be
\frac{d^2\phi}{dz^2}= \frac{dV}{d\phi}+ \beta(\phi) \frac{\rho(z)}{m_{\rm Pl}}
\ee
where the $z$ axis is perpendicular to the plates with $z=0$ on the bottom plate. The density is constant between the plates $\rho=\rho_b$ and inside them $\rho=\rho_c$. We also assume that $A(\phi)\sim 1$ for the variations of $\phi$ induced by $\rho$. This is satisfied for all the models we will study and comes from the Big Bang Nucleosynthesis (BBN) constraint on the variation of particle masses between BBN and now.

It is convenient to change variable from $\phi(z)$ to $a(z)$ where  $\phi(z)\equiv \phi (a(z))$ and $\phi(a)$ is given by (\ref{tom}). The Klein-Gordon equation becomes
\be
\frac{d^2 a}{dz^2} +\gamma(a) (\frac{da}{dz})^2= -\frac{a m^2(a)}{3}(1-\frac{a^3 H_\rho^2}{\Omega_{m0}H_0^2})
\ee
and we have defined the effective Hubble rate
\be
H_\rho^2= \frac{\rho}{3 m_{\rm Pl}^2}
\ee
which is constant inside and outside the plates. The function $\gamma(a)$ is given by
\be
\gamma (a)= \frac{d\ln\alpha}{da}
\ee
where
\be
\alpha (a)= \frac{\beta(a) }{a^4 m^2 (a)}.
\ee
Far enough inside the plates, $a(z)$ converges to a stationary value where the source term  vanishes for $a=a_c$ and we have
\be
\rho_c= \frac{\rho_0}{a_c^3}
\ee
where $\rho_0=3H_0^2 \Omega_{m0} m_{\rm Pl}^2$ is the matter density in the Universe now. For plates of common densities in the $\rho_c\sim 10\ \ {\rm g/cm^3}$ range, this corresponds to a very small $a_c\sim 0.5\ 10^{-10}$.
In this case we have that $a(z)\to a_c$ deep inside the plates.

It is useful to define the dimension-less functions $f$ and $g$ such that
\be
\beta(a)= \beta_0 f(a), \ m(a)= m_0 g(a)
\ee
where we normalise $f(0)=g(0)=1$. We also introduce the dimension-less  space variable
\be
u=m_0 z.
\ee
and the dimension-less field
\be
S(z)=\int^a_{a_c} da'\frac{f(a')}{a'^4 g^2(a')}
\ee
The corresponding effective potential
\be
V_S(S)= -\int^{a(S)}_{a_c} da \frac{f^2(a)}{3 a^7 g^2(a)}(1-\frac{a^3 H_\rho^2}{\Omega_{m0}H_0^2})
\ee
is such that the Klein-Gordon equation becomes
\be
\frac{d^2 S}{du^2}=\frac{\partial V_S}{\partial S}
\label{S}
\ee
Notice that this equation is independent of $\beta_0$, i.e. the field configuration does not depend on the value of the matter coupling now. This is a result which was already obtained for inverse power law chameleons and which is general for all tomographic models.
We can use this to integrate the Klein-Gordon.

\subsection{Bubbles}
We will assume that the plates are wide enough that $a(z)$ becomes constant deep inside them. This is tantamount to asking that the models we consider are such that large enough objects of high density screen the effects of the scalar field. In particular we shall require that the mass $m(a)$ becomes large enough inside the plates that the variation of $\phi$ occurs over a thin sheet close to the surface of the body. These conditions were already applied in the original  chameleon papers.

We can now integrate the Klein-Gordon equation. In the plates we have
\be
(\frac{dS}{du})^2= V_S^-(S).
\ee
where a $-$ sign signifies that the potential is defined with the density $\rho_c$ for $z<0$.
In between the plates we have
\be
(\frac{dS}{du})^2-(\frac{dS}{du})_s^2= 2(V_S^+(S)- V_S^+(S_s))
\ee
where $S_s=S(a(\vert z\vert =d))$ is the boundary value of $S$ and similarly for $(\frac{dS}{dz})_s$. The $+$ sign is to remind us that this potential is defined with the density $\rho_b$ between the plates.
Continuity implies that between the plates we have
\be
(\frac{dS}{du})_s^2= 2V_S^-(S_s)
\ee
and therefore the solution between the plate satisfies
\be
(\frac{dS}{du})^2= 2 (V^+_S(S)-\Delta V_s)
\ee
where $\Delta V_s= V_S^+(S_s)-V_S^-(S_s)$.
The solution $a(z)$ has a maximum at $z=d$. The resulting scalar configuration forms a bubble between the plates.  The profile of the bubble is determined for $0\le z\le d$ by the integral
\be
m_0z =\int_{S_s}^S \frac{dS'}{\sqrt{2(V^+_S(S)-\Delta V_s})}
\ee
and the extremal  value of $S$ is given by
\be
m_0d=\int_{S_s}^{S_d} \frac{dS'}{\sqrt{2(V^+_S(S)-\Delta V_s)}}
\ee
as a function of $S_c$ and $S_s$.
As there is an extremum at $z=d$ for $a_d=a(z=d)$, we have that $V^+_S(S_d)=\Delta V_s$ which implies that
\be
V^+_S(S)-V^+_S(S_d)= \int_{a(S)}^{a_d} da \frac{f^2(a)}{3 a^7 g^2(a)}(1-\frac{a^3 H_\rho^2}{\Omega_{m0}H_0^2}).
\ee
Between the plates  $\frac{H_\rho^2}{\Omega_{m0}H_0^2}=a_b^{-3}$ leading to
\be
V^+_S(S)-V^+_S(S_d)= \int_{a(S)}^{a_d} da \frac{f^2(a)}{3 a^7 g^2(a)}(1-\frac{a^3 }{a_b^3}).
\ee
The last term can be neglected as long as $a_d\ll a_b$. In this case, we have $S_d\ll S_b$ and the field deviates significantly from its value in the absence of both plates.
This determines the field profile completely and we get our final expression for the profile
\be
m_0(d-z)=  \int_{S}^{S_d} \frac{dS'}{\sqrt{2(V_S^+(S)-V_S^+(S_d))}}
\ee
where
\be
m_0d=  \int_{0}^{S_d} \frac{dS'}{\sqrt{2(V_S^+(S)-V_S^+(S_d))}}
\ee
These results are valid as long as the chamber is much larger than the range of the scalar field in the plates $m_c d\gg 1$, which guarantees that both plates are screened. We also assume that $m_b d \ll 1$ implying that the field is not sensitive to the exponential Yukawa fall off over distances greater than $m_b^{-1}$. We will apply these results to the models that we have presented.

\subsection{Power law models}

We focus on generalised power law models defined by
\be
m(a)= m_0 a^{-r},\ \beta=\beta_0 a^{-s}
\ee
which describe inverse power law chameleons,  the large curvature limit of $f(R)$ models for $a\ll 1$ and dilatons.
In this case we have that
\be
S(a)= \frac{1}{(2r-s-3)}(a^{2r-3-s}-a_c^{2r-3-s})
\ee
and the potential in between the plates
\be
V_S^+(a)-V_S^+(a_d)=\frac{1}{ 3(2r-6-s)}(a_d^{2r-6-2s}-a^{2r-6-2s})- \frac{1}{3(2r-3-2s)}(a^{2r-6-2s}(\frac{a}{a_b})^3-a_d^{2r-6-2s}(\frac{a_d}{a_b})^3)
\ee
We concentrate on cases where $a_d\ll a_b$, allowing us to neglect the terms coming from the density between the plates.
When the plates are screened, we have that $a(z)\gg a_c$ between the plates and therefore
\be
V_S^+(S)-V_S^+(S_d)=\frac{(2r-3-s)^{p}}{ 3(2r-6-s)}(S_d^{p}-S^{p})
\ee
where $p= \frac{2r-6-2s}{2r-3-s}$.
We restrict our attention to $2r-3-s>0$.

For all  power law models we have (see what follows)
\be
a_d\sim (m_0d)^{1/r}.
\ee
The conditions $a_c\ll a_d \ll a_b$ correspond to $ m_c^{-1} \ll d \ll m_b^{-1}$  which is the range of distances between the plates where the approximations we have used apply. When $d\gtrsim m_c^{-1}$, the field is constant between the plates and equal to $\phi_c$. When $d\gtrsim m_{b}^{-1}$, the influence of the two plates becomes negligible and the field converges to its constant value $\phi_b$.
We must now distinguish two cases
\subsubsection{Generalised chameleon models}
When $p<0$, we are in a situation similar to the case of inverse power law chameleons where $s=0$ and $r=3(n+2)/2(n+1)$. We find that
\be
S_d= K_p (m_0d)^{2/(2-p)}
\ee
and $K_p^{1-p/2}=\sqrt{2\frac{(2r-3-s)^{p}}{ 3\vert 2r-6-2s\vert I_p^2} }$ with $I_p=\int_0^1 \ \frac{dx}{\sqrt{x^p-1}}$. Notice that $S_d$ decreases when $d$ increases.
The bubble is defined by the integral
\be
I_p (1-\frac{z}{d})= \int_{\frac{S(z)}{S_d}}^1\frac{dx}{\sqrt{x^p-1}}.
\ee
Finally,
we can also express the field $S(z)$ when $z\ll d$ close to the first plate as
\be
S(z)= K_p((1-\frac{p}{2})I_p m_0 z)^{2/(2-p)}
\ee
which generalises the usual chameleon result and only depends on $m_0z$ and not on $d$ at all.

\subsubsection{Generalised f(R) models}

We consider models where $p>0$ generalising the large curvature $f(R)$ models where $s=0$ and $r=3(n+2)/2$. We find that the midpoint value between the plate is
\be
S_d= \tilde K_p (m_0d)^{2/(2-p)}
\ee
where $\tilde K_p^{1-p/2}=\sqrt{2\frac{(2r-3-s)^{p}}{ 3\vert 2r-6-2s\vert J_p^2} }$ with $J_p=\int_0^1 \ \frac{dx}{\sqrt{1-x^p}}$.
The bubble is defined by the integral
\be
J_p (1-\frac{z}{d})= \int_{\frac{S(z)}{S_d}}^1\frac{dx}{\sqrt{1-x^p}}
\ee
and the field $S(z)$ when $z\ll d$ close to the first plate is
\be
\frac{S(z)}{S_d}=  J_p \frac{z}{d}.
\ee
The field profile is linear in this case contrary to the generalised chameleon behaviour.
\subsubsection{Dilaton}
The dilaton can be described by the low density part of its potential where we have chosen $\phi_\star=0$ and
\be
r=3/2,\ s=-3
\ee
corresponding to a linear potential with $p=1$, i.e. the linear approximation an exponential potential in the corresponding range of field values.
This allows one to get exact expressions for the profile.
The midpoint value between the plate is
\be
S_d= \frac{(m_0d)^2}{6}.
\ee
The bubble is defined by the integral
\be
2(1-\frac{z}{d})= \int_{\frac{S(z)}{S_d}}^1\frac{dx}{\sqrt{1-x}}
\ee
and the field $S(z)$ becomes
\be
S(z)=S_d(1-(1-\frac{z}{d})^2)
\ee
and for small $z$
\be
\frac{S(z)}{S_d}=   \frac{2z}{d}
\ee
The field profile is also linear contrary to the generalised chameleon behaviour.

\subsection{Symmetron}
In the symmetron case, the field is at the origin deep inside the plates. Between the plates the field varies according to
 \be
 S(a)= S_\star \sqrt{1-(\frac{a_\star}{a})^3}
 \ee
 where
 \be
 S_\star= \frac{2}{3a_\star^3}
 \ee
 and the potential becomes
 \be
 V_S^+(S)-V_S^+(S_d)=\frac{1}{18a_\star^6}((1-\frac{S^2}{S_\star^2})^2 -(1-\frac{S_d^2}{S_\star^2})^2+2(1-\frac{S_b^2}{S_\star^2})(\frac{S^2}{S_\star^2}-\frac{S_d^2}{S_\star^2})))
 \ee
 corresponding to a quadratic potential close to the origin.
 Defining  $x=S/S_\star$ and $x_d=S_d/S_\star$ and the new variable
 \be
 x=\sqrt{1-(1-x_d^2)\cosh\theta}
 \ee
 where $\cosh \theta_d=1/(1-x_d^2)$, the maximal value $S_d$ is given by
 \be
 m_\star d= \int_0^{\theta_d}\frac{d\theta\sinh\theta }{\sqrt{(\sinh^2\theta + 2 \cosh\theta_b(1-\cosh\theta))(1-\frac{\cosh\theta}{\cosh\theta_d})}}
 \ee
 The symmetron has a non-vanishing profile in between the two plates only when $m_bd$ is larger than a critical value $m_b d_c$ obtained by taking $\theta_d$ to $0$. We find that
 \be
 m_b d_c= \frac{\pi}{\sqrt{2}}
 \ee
 where $m_b= m_\star \sqrt{1-\frac{a_\star^3}{a_b^3}}$ and we have assumed that $a_b>a_\star$, allowing the symmetron to probe the symmetry breaking part of its potential between the plates.
In the case when $m_b d\ll 1$ as in the case of cosmological symmetrons, we have $S_d=0$. Obviously in this case we have $S(z)=0$.

\section{Laboratory Tests}

\subsection{Casimir effect}
We will concentrate on the Casimir effect \cite{Lamoreaux:1996wh} induced by the presence of the scalar field coupled to the plates and having a bubble profile between the boundary plates.
Let us first rewrite the field equation inside and outside the plates
\be
\frac{d^2\phi}{dz^2}= \frac{\partial V_{eff}(\phi)}{\partial \phi}
\ee
from which we get the boundary value
\be
(\frac{d\phi}{dz})^2_s= 2(V_{eff}(\phi_d)-V_{eff}(\phi_s))=2(V_{eff}(\phi_s)-V_{eff}(\phi_c))
\ee
and using the explicit expression of $V_{eff}(\phi)$ we have
\be
A(\phi_s)= \frac{V(\phi_c)-V(\phi_d)+\rho_c A(\phi_c)-\rho_b A(\phi_d)}{\rho_c-\rho_b}
\ee
for the value of the field on the boundaries.

The Casimir force $F_\phi$  on one of the plates of surface area $A$ is simply obtained by integrating
\be
\frac{F_{\phi}}{A}=-\int_{d}^{D+d} dx  \rho_c\frac{dA(\phi)}{dx}
\ee
for a constant density plate of width $D$. We obtain the pressure
\be
\frac{F_{\phi}}{A}= -\rho_c(A(\phi_c)-A(\phi_s))
\ee
In the case where $\rho_c\gg \rho_b$, this expression simplifies and we get
\be
\frac{F_{\phi}}{A}= V(\phi_c)-V(\phi_d)+\rho_b(A(\phi_c)-A(\phi_d))
\ee
In the absence of a second plate, there is a vacuum pressure due to the scalar field where we replace $\phi_d\to \phi_b$ where $\phi_b$ is the minimum of the effective potential for a density $\rho=\rho_b$.
In a real experiment where the plates have a large but finite width, the vacuum pressure from the outside of the chamber on the plates would cancel leaving the plate in equilibrium if it were not for the presence of the second plate which offsets the pressure on the inner side of the plate. As a result, the vacuum pressure must be removed and the effective pressure felt by one plate is
\be
\frac{\Delta F_{\phi}}{A}= V_{eff}(\phi_b)-V_{eff}(\phi_d)
\label{cas}
\ee
corresponding to the difference between the effective potential in vacuum compared to the value it takes in between the plates.
This can be expressed as
\be
\frac{\Delta F_{\phi}}{A}=-27 \Omega_{m0}^2\beta_0^2 \frac{H_0^4 m_{\rm Pl}^2}{m_0^2}\int_{a_d}^{a_b} da \frac{f^2(a)}{ a^7 g^2(a)}(1-\frac{a^3 }{a_b^3})
\ee
where $\rho_b=\frac{\rho_0}{a_b^3}$. This proves that the scalar field adds an extra attracting pressure between the plate as the integrand is always positive. It is convenient to
rewrite this expression in terms of $V_S(S)$:
\be
\frac{\Delta F_{\phi}}{A}=-81 \Omega_{m0}^2\beta_0^2 \frac{H_0^4 m_{\rm Pl}^2}{m_0^2}(V_S^+(S_d)-V_S^+(S_b)).
\ee
The value of $S_d$ depends on the masses $m_c$ and $m_b$. When $m_cd\gtrsim 1$ and $m_b d\ll 1$, the field has a non trivial profile between the plates and we have calculated $S_d$ in the previous section. When $m_cd \lesssim 1$, the field is constant between the plates and $S_d=S_c$. Finally when the plates are not screened and $m_cD\lesssim 1$ where $D$ is the width of the plates, we have $S_d= S_b$ and no Casimir pressure is present.
We will use these results in the next section.

\subsection{The Eotwash experiment}

The search for the presence of new interactions by the Eotwash experiment \cite{Adelberger:2003zx} involves two plates separated by a distance $D$ in which holes of radii $r_h$ have been drilled regularly on a circle. The two plates rotate with respect to each other. The gravitational and scalar interactions induce a torque on the plates which depends on the potential energy of the configuration. The potential energy is obtained by calculation the amount of work required to approach one plate from infinity \cite{Brax:2008hh,Upadhye:2012qu}. Defining by $A(\theta)$ the surface area of the two plates which face each other (this is not the whole surface area because of the presence of the holes), a good approximation to the torque expressed as the derivative of the potential energy of the configuration with respect to the rotation angle $\theta$ is  given by
\be
T \sim a_\theta \int_D^{d_{\rm max}} dx (\frac{\Delta F_{\phi}}{A}(x)).
\ee
where $a_\theta=\frac{dA}{d\theta}$ depends on the experiment.
When the Casimir pressure due to the scalar field decreases fast enough with $d$, the upper bound $d_{\rm max}$ can be taken to be infinite. When this is not the case, the upper bound is the maximal distance below which the scalar force is not suppressed by the Yukawa fall-off. We will discuss the value of $d_{\rm max}$ for the different models that we have considered in the following section.
\subsection{Neutron energy levels}

Neutrons in empty space between two  mirrors have quantized energy levels in the terrestrial gravitational field \cite{Nesvizhevsky:2003ww,Brax:2011hb}. The scalar field induced a shift in the energy levels of the neutron due to the change in the potential energy
\be
V(z)=m_n gz + m_n (A(\phi (z))-1)
\ee
close to the lower mirror. The correction term is given by
\be
\delta V(z)=  9\Omega_{m0} \beta_0 m_n \frac{H_0^2}{m_0^2} \int_{a_c}^{a(S)} da \frac{f^2 (a) }{a^4 g^2(a)}
\ee
where $a(S)$ depends on $z$. This leads to a shift in the energy levels given by
\be
\delta E_n= <\psi_n\vert \delta V(z)\vert \psi_n>
\ee
where $\vert \psi_n>$ is the n-th Airy level of the neutron. We will evaluate this shift in the next section.

\section{Application to Models}

\subsection{Casimir effect}
We can now use the results of the previous section to calculate the effect of the scalar field, and its Casimir energy. We focus on the case where the plates are screened as in the absence of screening, no Casimir pressure is generated by the scalar field.
In the case of power models with $p<0$, the field value in the presence of the plates is much smaller than the one in their absence $S_d\ll S_b$ and we get
\be
V_S^+(S_d)-V^+(S_b)=\frac{(2r-3-s)^{p}}{ 3\vert 2r-6-2s\vert } K_p^p (m_0 d)^{2p/(2-p)}
\ee
corresponding to a Casimir Pressure
\be
\frac{\Delta F_{\phi}}{A}=\Lambda^4 (\frac{\sqrt{2p^2}}{B(\frac{1}{2},\frac{1}{2}-\frac{1}{p})} \Lambda d)^{\frac{2p}{2-p}}
\label{F}
\ee
where $B(.,.)$ is the Euler $B$ function. This generalises the inverse power chameleon case where $p=-n$ \cite{Brax:2007vm}. Notice that the Casimir pressure only depends on the scale $\Lambda$ and the distance $d$. When $\Lambda$ is taken to be the dark energy scale, this Casimir pressure is within reach of the next generation of Casimir experiments. We will present new forecasts at the end the paper.

For power law models with $p>0$, the contribution of $S_b=a_b^{2r-3-s}/(2r-3-s) $ cannot be neglected anymore.  In this case, we find a Casimir pressure
\be
\frac{\Delta F_{\phi}}{A}=27 \Omega_{m0}^2\beta_0^2 \frac{H_0^4 m_{\rm Pl}^2}{m_0^2}\frac{(2r-3-s)^{p}}{ \vert 2r-6-2s\vert } (\tilde K_p^p (m_0 d)^{2p/(2-p)}-\frac{3S_b^p}{2r-3-2s}).
\ee
We define the distance $d_\star$ where the $a_{d_\star}=a_b$, i.e. when the profile is that the scalar field begins to feel the effect of the matter density between the plates and its resulting suppression effect:
\be
(m_0d_\star)^{2p/2-p}=\frac{3 S_b^p}{(2r-3-s)\tilde K_p^p}.
\ee
We then find that
\be
\frac{\Delta F_{\phi}}{A}=81 \Omega_{m0}^2\beta_0^2 \frac{H_0^4 m_{\rm Pl}^2}{m_0^2}\frac{(2r-3-s)^{p}S_b^{p}}{ \vert 2r-6-2s\vert } ((\frac{d}{d_\star})^p-1)
\ee
As long as $d\ll d_\star$, the distance dependence becomes negligible and the pressure constant
\be
\frac{\Delta F_{\phi}}{A}=-81 \Omega_{m0}^2\beta_0^2 \frac{H_0^4 m_{\rm Pl}^2}{m_0^2}\frac{a_b^{2r-6-2s}}{ \vert 2r-6-2s\vert (2r-3-2s)}
\ee
as $p<2$ for all the power law models with $r>0$.

The symmetron case  leads to a constant pressure too. As long as $m_b d\ll 1$, we have that
\be
S_d=0
\ee
and the Casimir pressure is given by a constant
\be
\frac{\Delta F_{\phi}}{A}=-\frac{9}{2} \Omega_{m0}^2\beta_0^2  \frac{H_0^4 m_{\rm Pl}^2}{a_\star ^6 m_0^2}
\ee
We can rewrite this result as
\be
\frac{\Delta F_{\phi}}{A}=-\frac{\mu^4}{4\lambda}
\ee
which is the height of the symmetron potential.

\subsection{Gravitational experiment}

We can use the previous result on the Casimir pressure to infer the torque on the rotating plates in the Eotwash experiment in the screened case. Let us first focus on power law models. When $p<0$,
the Casimir pressure falls off at infinity and two cases must be envisaged. When $p<-2$, the fall is fast enough that no dependence on $d_{\rm max}$ is of relevance and
\be
T_\theta= a_\theta  \frac{2-p}{p+2}(\frac{\sqrt{2p^2}}{B(\frac{1}{2},\frac{1}{2}-\frac{1}{p})})^{\frac{2p}{2-p}}\Lambda^3 (\Lambda d)^{(p+2)/(2-p)}
\ee
When $-2<p<0$, the torque is sensitive to the long distance behaviour of the Casimir pressure which becomes negligible when $d= d_\star$ where $a_{d_\star}= a_b$, i.e. we take $d_{\rm max}=d_\star$, implying that
\be
T_\theta= a_\theta  \frac{2-p}{p+2}(\frac{\sqrt{2p^2}}{B(\frac{1}{2},\frac{1}{2}-\frac{1}{p})})^{\frac{2p}{2-p}}\Lambda^3[ (\Lambda d)^{(p+2)/(2-p)}-(\Lambda d_\star)^{(p+2)/(2-p)}].
\ee
which is independent of $d$ as long as $d\ll d_\star$ where we have here
\be
(m_0d_\star)^{2p/(2-p)}=\frac{3 S_b^p}{(2r-3-s) K_p^p}.
\ee
Notice that the torque depends on the combinations $\Lambda d$ and $\Lambda d_\star$, i.e. it probes distances of the order of the inverse dark energy scale which is about $\Lambda^{-1}\sim 82 \ \mu{\rm m}$. Its order of magnitude is then around $a_\theta\Lambda^3$ which is very close to the bound found by Eot-wash.

For power law models with $p>0$, the situation is similar to the case $-2<p<0$ where the long range nature of the Casimir force is crucial
\be
T_\theta= a_\theta \frac{81}{2r-6-2s}\frac{\Omega_{m0}^2 \beta_0^2 H_0^4 m_{\rm Pl}^2}{m_0^2} S_b^p d_\star [\frac{2-p}{p+2}(1- (\frac{d}{d_\star})^{(p+2)/(2-p)})-(1-\frac{d}{d_\star})].
\ee
As $0<p<2$ for the models that we consider, this simplifies to
\be
T_\theta= -a_\theta \frac{162}{2r-6-2s}\frac{p}{p+2}\frac{\Omega_{m0}^2 \beta_0^2 H_0^4 m_{\rm Pl}^2}{m_0^2} S_b^p d_\star
\ee
which is a function of $m_0$ and $\beta_0$.

For the symmetrons, the Casimir force is independent of the distance as long as the field vanishes between the plates. This is true as long as $d< d_c= \frac{\pi}{\sqrt 2 m_b}$, hence the torque is given by
\be
T_\theta=- a_\theta \frac{\mu^4 d_c}{4\lambda}
\ee
which depends on $\mu$ and $\lambda$. The dependence on the coupling strength $\beta_\star$ only appears when the electrostatic shielding between the plates is taken into account.

\begin{figure}
\centering
\includegraphics[width=0.50\linewidth]{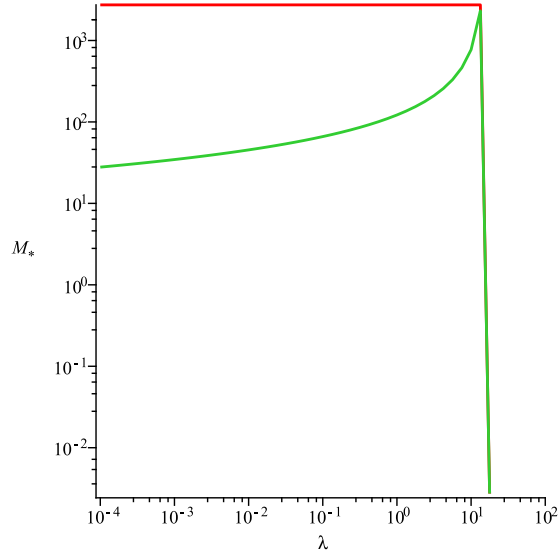}
\caption{The phase diagram of symmetron models for $\mu=\Lambda=2.4\ 10^{-3}$ eV as an example where $d<d_c$ and the field vanishes between the plates. The Eot-wash experiment is sensitive to symmetrons for small values of $\lambda$ to the left of the vertical line. For values of $M_\star$ larger than the top horizontal line, the symmetron is in its vacuum phase and no constraints apply.  Below the bottom curve the symmetron is not excluded while it is excluded between the bottom curve and the  horizontal line.    }
\end{figure}
\subsection{Neutron energy levels}

For power law models we have that
\be
\delta V(z)=  9\Omega_{m0} \beta_0 m_n \frac{H_0^2}{m_0^2} \frac{1}{2r-3-2s}((2r-3-s) S(z))^{p}.
\ee
When $p<0$, we generalise the inverse power law chameleons and we find
\be
\delta V(z)= \frac{\beta_0 m_n}{m_{\rm Pl}} \Lambda (\frac{2-p}{\sqrt{2}} \Lambda z)^{2/(2-p)}
\ee
which has been thoroughly studied\cite{Brax:2013cfa}.

Focusing on the models with  $p>0$ as they differ from the behaviour of inverse power law chameleons.
In this case we obtain that
\be
\delta V(z)=  9\Omega_{m0} \beta_0 m_n \frac{H_0^2}{m_0^2} \frac{1}{2r-3-2s}((2r-3-s) J_p S_d)^{\frac{2r-3-2s}{2r-3-s}}(\frac{z}{d})^{\frac{2r-3-2s}{2r-3-s}}
\ee
The shift in the energy levels is then given by
\be
\delta E_n= 3\Omega_{m0} \beta_0 m_n \frac{H_0^2}{m_0^2} \frac{1}{2r-3-2s}\alpha_{n,r}((2r-3-s) J_p S_d)^{\frac{2r-3-2s}{2r-3-s}}(\frac{z_0}{d})^{\frac{2r-3-2s}{2r-3-s}}
\ee
where the numbers
\be
\alpha_{n,r}=<\psi_n\vert (\frac{z}{z_0})^{\frac{2r-3-2s}{2r-3-s}}\vert \psi_n>
\ee
are of order one, and $z_0= \left(\frac{1}{2m_{n}^2 g}\right)^{1/3}$. In practice, the distance between the plates is adjusted to select different energy levels and therefore $z_0\sim d$.
The correction to the energy levels has a dependence on
\be
\delta E_n\sim  \beta_0 m_n \frac{H_0^2}{m_0^2} (m_0d)^{(2r-3-2s)/r}
\label{ree}
\ee
Contrary to $p<0$, the result depends on $d$, i.e. on the details of the experimental setup. It also depends on the ratio of  the distance between the two plates and the size of the present horizon. For the present day sensitivities at the $10^{-14}$ eV, the deviation $\delta E_n$ is not observable for both $f(R)$ models and dilatons.
Finally, for cosmological symmetrons we have that $S(z)=0$ and no deviation of the energy levels is expected.
\section{Field Theoretic Consistency}
So far we have treated the field theoretical models as classical field theories. Most of the theories we have discussed have unusual features such as inverse power law potentials and non integer powers of the field. In this section we will describe their properties using the language of effective field theories when they are embedded in an environment with a uniform energy density. In this case, the tomographic field theories have a well-defined minimum of the effective potential around which one can expand the potential in perturbation. This will allow us to discuss their validity and the quantum corrections which can be easily calculated at the one loop order.

\subsection{Tomographic models as effective field theories}

The tomographic models in the presence of a constant density environment are field theories with a potential described by
\be
V_{\rm eff}= -\frac{27 \Omega_{m0^2} \beta_0^2 H_0^4}{m_0^2}\int_{a_c}^{a(\phi)} da \frac{f^2(a)}{3 a^7 g^2(a)}(1-\frac{a^3 H_\rho^2}{\Omega_{m0}H_0^2})
\ee
where $a(\phi)$ has to be computed using the mapping
\be
\phi(a)=\frac{9\Omega_{m0} \beta_0 H_0^2 m_{\rm Pl}}{m_0^2} \int_{a_c}^{a(\phi)} da \frac{f(a)}{3 a^4 g^2(a)}.
\ee
These theories can be expanded around a background field value $\phi$ corresponding to a value of $a=\bar a$ in an infinite series which defines the tree level Lagrangian
of an effective theory
\be
{\cal L}= \frac{1}{2} (\partial \delta \phi)^2 + \sum_{i=0}^\infty \frac{\lambda_i}{i!} \delta \phi^i
\ee
 where we have
$
\lambda_i= \frac{d^i V_{\rm eff}}{d\phi^p}\vert_{\phi=\bar\phi}.
$
This can be easily reexpressed as
$
\lambda_i= (\frac{m_0^2}{9\Omega_{m0}\beta_0 H_0^2 m_{\rm Pl}})^{i-2} \tilde \lambda_i
$
where we identify the dimension-less coupling
$
\tilde \lambda_i= \frac{d^i V_S}{dS^i}\vert_{S=\bar S}
$
and
$
\bar S= \int_{a_c}^{\bar a} da \frac{f(a)}{3 a^4 g^2(a)}.
$
The infinite series can be rewritten as
\be
{\cal L}= \frac{1}{2} (\partial \delta \phi)^2 + \sum_{i=0}^\infty (\frac{\tilde \lambda_i}{i!}) (\frac{m_0^6}{81\Omega_{m0}^2 \beta_0^2 H_0^2 m_{\rm Pl}^2})\frac{\delta \phi^i}{\tilde \Lambda_0^{i-4}}
\ee
with
$
\tilde \Lambda_0=\frac{9 \Omega_{m0}\beta_0 H_0^2 m_{\rm Pl}}{m_0^2}.
$
The terms with $i<4$ are the relevant terms of the effective field theory defined by the infinite series. The term $i=4$ is the marginal interaction with a dimension-less coupling constant while the $i>4$ terms are the
irrelevant interactions at low energy which are non-renormalisable operators. This will tell us when we can truncate the infinite series and keep only the terms up to $i=4$ at low energy, below a cut-off scale that we will determine. It will also tell us when the perturbative expansion makes sense and no strong coupling issue arises.

\begin{figure}
\centering
\includegraphics[width=0.50\linewidth]{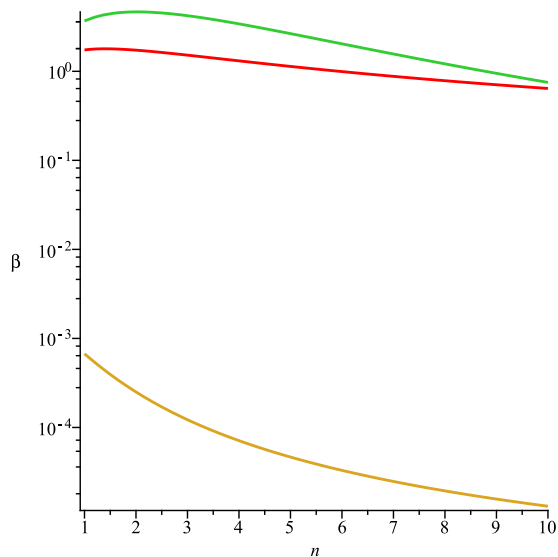}
\caption{Chameleons are such that the plates are not screened  below the bottom curve (brown) and quantum effects are strong for values of the coupling larger than the top curve (green). The  middle curve (red) is the limit below which the field is constant between the plates. }
\end{figure}

Practically we have
\be
\tilde \lambda_1= -\frac{1}{3} f(a) (a^{-3} -a_c^{-3})
\ee
 and
 \be
\tilde \lambda_i = \frac{a^4 g(a)}{f(a)} \frac{d\tilde \lambda_{i-1}}{da}
\ee
recursively.
Explicitly we find that the second coupling is
\be
\tilde \lambda_2= g(a)^2 - \frac{d\ln f}{da} g^2(a) a^4(a^{-3} -a_c^{-3})
\ee
We are considering the effective field theory when the matter density is $\rho_c$ and we are expanding around  the vacuum value $a=a_c$.
In this case $\tilde \lambda_2= g^2$. To go further, we shall assume that around $a_c$, the mass function has a power law dependence $g(a)=a^{-r}$ and the coupling function $f(a)=a^{-s}$.
In this case we find that
\be
\tilde \lambda_i \sim a^{-6r+6-2s} (a^{3-2r+s})^{i-4}
\ee
implying that the infinite series can be written as
\be
{\cal L}= \frac{1}{2} (\partial \delta \phi)^2 + \sum_{i=0}^\infty (\frac{ \kappa_i}{i!}) \frac{\delta \phi^i}{\Lambda_0^{i-4}}
\ee
where the coupling constants are obtained to be
\be
\kappa_i \sim \frac{m^6(\bar \phi) m_{\rm pl}^2}{\beta^2(\bar \phi)\rho_c^2}
\ee
and the cut-off is
\be
\Lambda_0= \frac{\beta(\bar\phi)\rho_c}{m^2(\bar\phi)m_{\rm Pl}}.
\ee
This is the final form of the effective action for tomographic models. First of all, we find that the tomographic models viewed as effective field theories are only defined at low energy below
the cut-off $\Lambda_0$. The smallest cut-off scales is obtained in dense matter  where $\Lambda_0 \sim a_c^{2r-3-s}\frac{\beta_0 H_0^2}{m_0^2} m_{\rm pl}$. For cosmologically relevant models where $m_0\gtrsim 10^3 H_0$, this is a very small scale much smaller  than the scale of the standard model of particle physics. Hence all the tomographic models in dense matter have a very low cut-off scale. Similarly, the tomographic models can be strongly coupled in dense matter when
\be
m(\bar \phi) \gtrsim (\frac{\beta(\bar\phi)\rho_c}{m_{\rm Pl}})^{1/3}.
\ee
In this case, the effective field theory is not well-defined and one cannot consider that tomographic models describe the behaviour of massive particles with self-interactions at an energy below the cut-off scale $\Lambda_0$.
This does not imply that the tomographic models are not well defined themselves. It simply means that one cannot describe them as an effective field theory around a constant background field $\bar\phi$. For these cases, perturbation theory fails.

\subsection{Quantum corrections}

Quantum corrections and the quantum stability of the tomographic models are very important to guarantee that the results we have obtained at tree level for the laboratory experiments such as the Casimir effect stand when quantum effects are taken into account. The quantum effects of the scalar field $\delta\phi$ can be calculated when the effective field theory is not strongly coupled and when the scalar field has a mass below the cut-off scale
$\Lambda_0$. Fortunately, the latter is equivalent here to requiring that the model is not strongly coupled. In this case, the one loop correction to the potential is
\be
\delta V= \frac{m^4(\bar\phi)}{32\pi^2} \ln \frac{m(\bar\phi)}{\Lambda_0}
\ee
Around the vacuum value $a=a_c$, the effective potential
\be
V_{\rm eff}(\bar\phi)= -\frac{27 \Omega_{m0^2} \beta_0^2 H_0^4}{m_0^2}\int_{a_c}^{a(\bar \phi)} da \frac{f^2(a)}{3 a^7 g^2(a)}(1-\frac{a^3 H_\rho^2}{\Omega_{m0}H_0^2})+\frac{m^4(\bar\phi)}{32\pi^2} \ln \frac{m(\bar\phi)}{\Lambda_0}
\ee
has a new minimum which is shifted $a_c\to a_c+\delta a$ where
\be
\frac{\delta a}{a_c}=\frac{r}{48\pi^2}\frac{m^6_c m^2_{\rm pl}}{\beta_c^2\rho_c^2} \ln\frac{m_c}{\Lambda_0}
\label{cor}
\ee
The quantum corrections are negligible when \cite{Upadhye:2012vh}
\be
m_c\le (\frac{6\pi \sqrt{2}  \beta_c\rho_c}{\sqrt{r} m_{\rm Pl}})^{1/3}
\ee
which is similar to the requirement that the theory must not be strongly coupled. Hence we have found that tomographic models which are not strongly coupled at tree level do not suffer from any quantum
instability.

The quantum corrections are larger in dense media. In the cosmological vacuum, the absence of quantum correction is guaranteed when
\be
\frac{m_0}{H_0}\lesssim  (\frac{\beta_0 m_{\rm pl}}{H_0})^{1/3}
\ee
For theories with $\beta_0 \gtrsim 1$, the right hand side is of order $10^{20}$, hence for cosmological models where $m_0/H_0\gtrsim 10^3$ the quantum corrections are always negligible on cosmological scales. For dense media, quantum corrections can play a major role.

\subsection{Strong coupling phase}

The failure to calculate quantum corrections and the strong coupling issue can be resolved in certain cases when a constant background field $\bar\phi$ is not an appropriate description of the vacuum structure
of the model. This is in particular the case for inverse power law chameleons. In \cite{Mota:2006ed,Mota:2006fz,Brax:2013cfa}, it was shown that for large values of $\beta_0$, the nuclei in each atom of the dense medium become screened when
\be
\phi_c\le 2 \beta_0 m_{\rm Pl} \Phi_n
\ee
where $\Phi_n= \frac{m_n}{8\pi m_{\rm Pl^2} R_n}$ whilst $m_n$ and $R_n$ are the mass and the radius of the nucleus. In this case, the homogeneous solution where $\phi=\phi_c$ inside the dense body is not valid anymore.
The scalar field becomes inhomogeneous and forms bubbles centered at each atom and similar to the bubble solution between two plates but on atomic scales. Outside a radius $R_\star \sim \bar D (\frac{R_n}{\bar D})^{(n+1)/(2n+1)}$ the solution  for $R_\star \ll r \ll R$  grows like a bubble in $\phi(r) \sim \Lambda (\frac{n+2}{\sqrt 2} \Lambda r)^{2/(n+2)}$ before reaching a maximum
\be
\phi_D= (\frac{\sqrt{2} \Lambda \bar D}{I_n})^{2/(n+2)}
\ee
where $2\bar D$ is the interatomic distance. Apart from the steep increase around each atom, the solution is of average $\phi_D$ which is very different from the homogeneous solution. In fact we have the relation between $\phi_D$ and  the maximal value between  the boundary plates of the experiments that we have considered
\be
\phi_D= (\frac{\bar D}{d})^{2/(n+2)} \phi_d.
\ee
implying that $\phi_D\ll \phi_d$. This boundary value is small enough to guarantee the existence of the bubble solution between the plates\footnote{For even larger values of $\beta_0$, the cloud of electrons in the plate serves as the source for the scalar field which becomes homogeneous again. For chameleons, this happens for very large couplings which are already excluded \cite{Upadhye:2012ar}.}.

For such values of $\beta_0$, the quantum corrections to the chameleon potential inside the nuclei can be  large and therefore not calculable. In this case this implies that the bubble structure between atoms may be  destroyed by those quantum corrections inside the nucleus if the quantum corrected minimum inside the nuclei is much larger than its tree level value. Unfortunately, the behaviour of inverse power law models in nuclear matter goes beyond the domain of validity of such models which we have assumed to be valid for densities smaller or equal to the ones during BBN, i.e. a few $\rm{g/cm^3}$. At much larger densities, the models have to altered to guarantee that the quantum corrections are negligible. This can be achieved by appropriately modifying the matter $m(a)$ and coupling function $\beta(a)$ at tree level for $a\lesssim a_{BBN}$. For instance an interpolation to a symmetron-like behaviour at high density
\be
m(a)=m_0 a^{-r}\sqrt{1-(\frac{a_\star}{a})^3},\ \ \beta(a)=\beta_0 \sqrt{1-(\frac{a_\star}{a})^3}
\ee
with $a_\star < a_{BBN}$ and
where the field would be stuck to a vanishing value in the regime $a<a_\star$ would annul  the effects of the quantum corrections.  The details are left for future work.

\subsection{Application to laboratory experiments}

For the laboratory experiments that we have considered where two dense plates are separated by a gas, the quantum corrections are  relevant for models of the generalised chameleon type as inside the plates
the quantum corrections can be large. When the quantum corrections are not negligible, the minimum of the effective potential can be shifted by a large amount or even disappear. On the other hand, the field equations between the plates are not affected by the quantum corrections. Hence the only role played by the potentially large quantum corrections is to modify the boundary values of the field and its first derivative on the plates.

The derivation of the Casimir pressure (\ref{cas}) is not affected much, implying that the only effect of the quantum corrections is to shift the boundary value $\phi_s$ and therefore to modify the value $\phi_d$ between the plates. Using (\ref{cor}) as an order of magnitude estimate of the value $a_q$ of the new minimum
\be
a_q\sim a_c \frac{m^6_c m_{\rm Pl}^2}{\beta_c^2 \rho_c^2}.
\ee
Therefore the structure of the bubble between the plate is preserved as long as $a_q\ll a_d$. When this is not the case, the boundary value is affected too much to guarantee that a bubble solution still exists between the plates and therefore the classical results are largely affected.

At  higher values of the coupling, chameleon models are described by an inhomogeneous solution at the atomic level. Again this solution is highly sensitive to the quantum stability of chameleons in nuclear matter. As long as the coupling in this dense environment is small enough, which requires us to modify the models for density higher than the BBN ones, the bubbles formed at the atomic level are quantum stable and the bubble solution between the boundary plates is preserved. In this case, the experiments probing the large coupling limit of chameleon models are immune from quantum corrections.

\section{Constraints and Forecast}
\subsection{Constraints}

The most stringent experimental constraint on the intrinsic value of the Casimir pressure has been obtained with a distance $2d=746$ nm between the two boundary plates and reads $\vert \frac{\Delta F_{\phi}}{A}\vert \le 0.35$ mPa where
we have $1\rm { mPa}= 1.44 \ 10^6 \Lambda^4$ \cite{Decca:2007yb}. The experiment has been performed with a pressure of $10^{-4}$ Torr between the plates of width $D=0.5 $ cm corresponding to $\rho_b=6.7\ 10^{-28}\ {\rm GeV}^4$ and $a_b\sim 3\cdot \ 10^{-7}$. The plate density is of the order of  $\rho_c=10\ {\rm g.cm^{-3}}$. For the 2006 Eot-wash experiment \cite{Kapner:2006si}, we consider the bound obtained
 for a separation between the plates of $2d=55\mu{\rm m}$ is
\be
\vert T \vert \le a_\theta \Lambda_T^3
\ee
where $\Lambda_T= 0.35 \Lambda$ \cite{Brax:2008hh}. The pressure was lower than in the Casimir experiment corresponding $10^{-6}$ T and $a_b\sim 1.4\ 10^{-6}$.
We must also modify the torque that we have calculated in order to take into account the effects of a thin electrostatic shielding sheet of width $d_s=10\mu {\rm m}$ between the plates in the Eotwash experiment. This reduces the observed torque which becomes
\be
T_{obs}=e^{-m_c d_s} T_\theta
\ee
When the mass in dense media is very large, this imposes a strong reduction of the signal.

We combine these bounds as an illustration for $f(R)$ models in the large curvature limit, dilatons, symmetrons and inverse power law chameleons. For $f(R)$ models, the Casimir bound on $m_0/H_0\gtrsim 10^{-12}$ for $n\gtrsim 0.01$ is irrelevant compared to the usual bound from solar system tests $m_0/H_0\gtrsim 10^3$. For values of $m_0$ satisfying the solar system bound, the effect of the electrostatic shielding is so large that $f(R)$ models cannot be tested by these experiments as $m_c d_s\gg 1$.
For dilatons with $\beta_0\sim 2.7$, the electrostatic shielding is  efficient when $m_0/H_0 \gtrsim 10^{16}$, implying that no constraint can be obtained for such large masses. For masses $m_0/H_0\gtrsim 10^{16}$, quantum corrections become large in the plates. The plates are screened provided $m_0/H_0\gtrsim 2\cdot 10^{15}$. Below this value, the field between the plates is equal to the one in the plates. This  does not change the leading order expression for the torque and we find that the Eot-wash experiment requires that $m_0/H_0\gtrsim 55$. This is less stringent than the solar system tests coming from the Lunar Ranging experiment $m_0/H_0\gtrsim 34500$.

\begin{figure}
\centering
\includegraphics[width=0.50\linewidth]{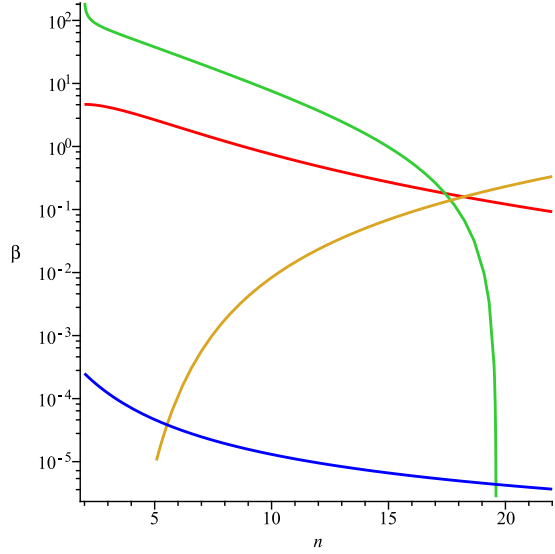}
\caption{ The top curve (green)  represents the Eot-wash limit. It is in the quantum corrected part of the phase diagram of chameleons which corresponds to $\beta$ being larger than values on the middle curve (red) where quantum effects start being relevant. The lower curve (blue) represents the limit below which the plates are not screened and the torque vanishes. This implies that chameleons are not excluded for very low couplings. The ascending curve (brown) is the Eot-wash limit when the field between the plates is constant and equal to the values inside the plates for $m_c d\lesssim 1$. The triangular region between these three curves (red,blue,brown) is excluded by the Eotwash experiment. For large values of $n\gtrsim 20$, no constraints from Eot-wash apply. Notice that the top part of the exclusion zone is in the quantum corrected region of the parameter space and therefore cannot be trusted. }
\end{figure}

The symmetrons are strongly constrained by the Eot-wash experiment. Let us define
\be
M_\star^2= \frac{\phi_\star m_{\rm pl}}{\beta_\star},
\ee
the mass of the scalar field when the density is much larger than the critical density $\rho_\star$ is given by
\be
m(a)=\mu (\frac{a_\star}{a_c})^{3/2}
\ee
The Eot-wash bound can be expressed as
\be
\Lambda_T^3 \ge \frac{\pi}{4}\frac{\mu^3}{\lambda} e^{-(\frac{a_\star}{a_c})^{3/2}\mu d_s}
\ee
which can be written as
\be
M_\star\le \frac{\sqrt{\rho_c} d_s}{\ln(\frac{\pi\mu^3}{4\lambda \Lambda_T^3})}
\ee
as long as $4\lambda \Lambda_T^3 \le \pi \mu^3$.
The torque calculation that we have presented  applies only when $a_\star\ge a_c$ where
\be
a_\star=(\frac{\rho_0}{\mu^2 M_\star^2})^{1/3}.
\ee
For large values of $M_\star$, the symmetron is always in its vacuum phase and there is no torque between the plates.
The combined constraints can be seen in Figure 1 for $\mu=\Lambda$. For large values of $\lambda$, no constraints are obtained while at small $\lambda$, values of $M_\star$ are bounded from above.
This part of the parameter space can be tested by laboratory experiments simply because $a_\star \le a_b$. Cosmologically interesting models where $\mu \sim H_0$ and $a_\star\ge 10^{-1}$ corresponding to a phase transition
in the recent Universe cannot be probed by current experiments as $a_b$ is too small. For these models, the field value in all the laboratory environment is equal to zero, implying that no Casimir effect or torque is present. When $a_b >a_\star$, the parameter space is constrained by
\be
\beta_\star^2 \le \frac{4}{\pi} a_\star^6 \frac{m_{\rm Pl}^2 \Lambda_T^3 \mu^3}{\rho_0^2} e^{(\frac{a_\star}{a_c})^{3/2}\mu d_s}
\ee
and for $\mu\sim H_0$ and small values of $a_\star$, the Eot-wash experiment implies a tight constraint on $\beta_\star\lesssim 10^{-18}$.

\begin{figure}
\centering
\includegraphics[width=0.50\linewidth]{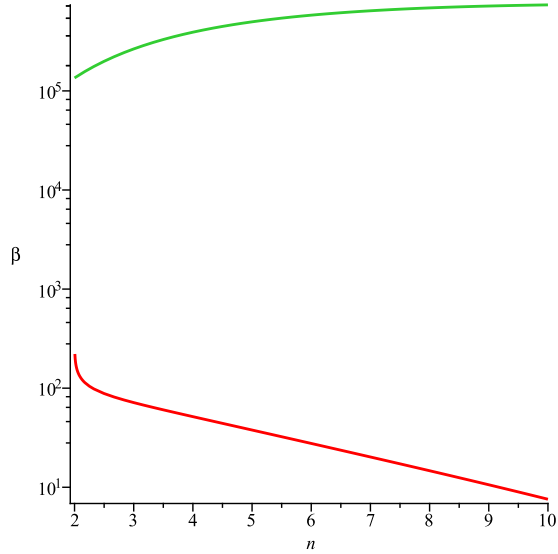}
\caption{For large values of $\beta_0$ higher than the top curve (green), the chameleons are in their strongly coupled phase and the quantum corrections which are present for values larger than the lower curve (red) are not important anymore.
}
\end{figure}

Finally, for inverse power law chameleons the phenomenology is richer. We focus here on the Eot-wash experiment. At small coupling $\beta_0$, the boundary plates are not screened. The Eot-wash bound is satisfied for values of $\beta_0$ larger than a lower bound which depends on $n$. For $n=1$, we must have $\beta_0\gtrsim 135$. On the other hand, for such values of $\beta_0$ the quantum corrections are not under control in the plates and the bubble solution between the plates is not calculable anymore, see Figure 2. In fact as we show in Figure 3 the quantum bound is always lower than the Eot-wash upper bound for $n\lesssim 20$ implying that present day experimental results are not compatible with the absence of quantum corrections in the part of the parameter space between the top (green) and middle (red) curves of Figure 3. For $n\gtrsim 20$, the Eot-wash result does not constrain chameleons due to the large suppression by the electrostatic shield. For larger values of $\beta_0\gtrsim 10^5$, chameleons are in their strong coupling phase where bubbles appear at the atomic scale. This guarantees that the bubble between the plates is present and the Eot-wash bound applies, i.e. the strong coupling models are allowed.
Finally, for low values of $\beta_0$, the plates are separated by a distance lower than the distance $d_{\rm screen}$ defined by $d_{\rm screen}= m_c^{-1}$ and the field between the plates becomes equal to $\phi_c$. This modifies the expression of the torque
\be
T_{obs}=a_\theta e^{-m_c d_s}[ \int_d^{d_{\rm screen}}(V(\phi_c)-V(\phi_b))dx + \int_{d_{\rm screen}}^{d_\star} \frac{\Delta F_\phi}{A} dx]
\ee
where the Casimir pressure $\frac{\Delta F_\phi}{A}$ is given by (\ref{F}). The first term goes to zero as $\beta_0\to 0$. On the other hand, the second term depends on $n$. For $n>2$, the integral is dominated by $x \sim d_{\rm screen}$ which increases as $\beta_0\to 0$ and therefore the integral goes to zero. In this case, the torque becomes smaller and smaller for $\beta_0\ll 1$ and chameleons with $n>2$ become unobservable. On the other hand for $n\le 2$, the integral is dominated by its upper bound and increases with it. The torque remains large for small $\beta_0$. This is valid as long as $m_cD\gtrsim 1$. When this is not the case anymore, the torque vanishes and the chameleons are not constrained by the Eot-wash experiment (see Figure 3).

\begin{figure}
\centering
\includegraphics[width=0.50\linewidth]{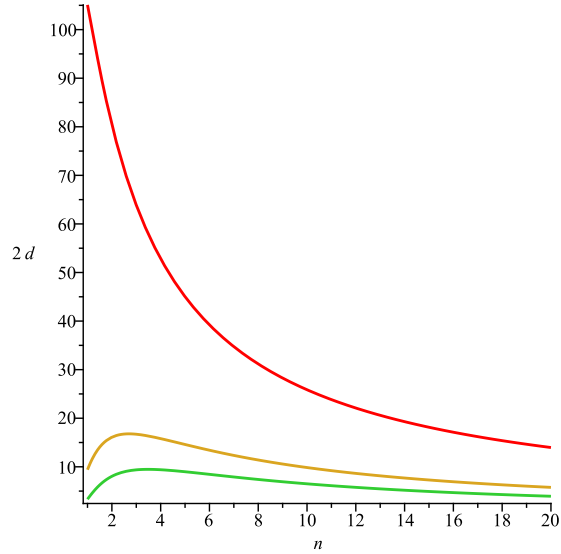}
\caption{The distance below which new Casimir experiments with plates of surface areas $1\ \rm {cm^2}$ would feel a scalar Casimir force of respectively from  top to bottom $0.1 $ pN, $0.5$ pN and $1$ pN. At distances $2d=10\ \mu{\rm m}$, the new Casimir experiments with a sensitivity of $0.1$ pN would cover most of the chameleons cases. }
\end{figure}

\subsection{Forecast}

 Casimir tests are particularly relevant for inverse power law chameleon models and their generalisations. In the regime where the boundary plates are screened, and in particular for very large values of $\beta$, the Casimir pressure due to the chameleons is independent of $\beta$. The next generation of Casimir experiments such as CANNEX in Amsterdam will test the electromagnetic Casimir effect at distances larger than $2d\gtrsim 10 \mu{\rm m}$, with a sensitivity which could reach $0.1$ pN for plates of surface area $A=1{\rm cm^2}$. At large coupling, the Casimir force between the plates due to the scalar field is independent of $\beta_0$ and depends only on $n$. In figure 4 we show, for such a setting, the distance between the plates for which the scalar Casimir force reaches the requires sensitivity. In particular, we find that at the $0.1$ pN level and below, most inverse chameleon models will be tested at a distance $2d=10\mu{\rm m}$. Even with a precision of $0.5$ pN, the parameter space of chameleons will be almost covered for reasonable values of $n$.

\section{Conclusion}

We have considered laboratory tests of tomographic models such as $f(R)$ in the large curvature regime, inverse power law chameleons, dilatons and symmetrons. We have given simple analytical expressions for the Casimir pressure due to the scalar, the torque between two rotating plates in the Eot-wash experiment, and the shift of the energy levels of the neutron in the terrestrial gravitational field. We have analysed in detail the behaviour of $f(R)$ models, inverse power law chameleons, dilatons and symmetrons in these experiments. We have also analysed the quantum corrections which can be particularly large for inverse power law chameleon models. We have shown that these quantum corrections are under control when the tomographic models are at small coupling in a perturbative expansion around the density dependent minimum of their effective potential. At strong coupling, and for models such as inverse chameleon models, the homogeneous approximation which has been used to calculate the quantum correction fails. The scalar field becomes inhomogeneous at the atomic level and as long as the coupling to nuclear matter is assumed to be vanishingly small, the quantum corrections are tamed at large coupling for matter densities lower than the BBN one. In this strong coupling phase, inverse power law chameleons are within reach of the next generation of Casimir experiments.

We would like to thank C. Burrage and A. Upadhye for comments and suggestions. We are grateful to A. Amalsi and R. Sedmik for discussions and sharing information about the CANNEX experiment, and to G. Pignol for lively discussions about the dilaton and the neutron experiments.
P.B.
acknowledges partial support from the European Union FP7 ITN
INVISIBLES (Marie Curie Actions, PITN- GA-2011- 289442) and from the Agence Nationale de la Recherche under contract ANR 2010
BLANC 0413 01. ACD acknowledges partial support from STFC under grants
ST/L000385/1 and ST/L000636/1.

\end{document}